# Negative thermal expansion induced suppression of wear in dry sliding friction

Aleksandr S. Grigoriev[*], Evgeny V. Shilko, Andrey I. Dmitriev, Sergey Yu. Tarasov

*Institute of Strength Physics and Materials Science SB RAS, Tomsk 634055, Russia*

Surface temperature is among crucial factors, which control wear during sliding dry contact. Using computer modeling, we study the possibility to achieve close to zero rate of surface wear during sliding friction of the special type of materials, which possess negative thermal expansion. The numerical simulations reveal two wear regimes for negative thermal expansion materials as dependent on the applied normal stress level. When the applied stress is lower than that of a critical level, a steady almost zero wear rate and nanorough surface are achieved during friction. Otherwise, wear rate is of the same order of magnitude as for positive thermal expansion materials. The critical stress value is analyzed as depending on the material's mechanical, thermophysical, and surface roughness characteristics.

Temperature is a fundamental factor to control friction and wear in a solid/solid contact at all scales. The locality of energy dissipation in rubbing between rough surfaces results in the fact that the mean subsurface layer temperature may be much higher than that of the bulk of the body and moreover, the so-called flash can be higher 1000K [1-7]. Temperature dependencies of mechanical characteristics, acceleration of phase transformations, and mechanochemical reactions, activation of special thermolubrication mechanisms as well as some other phenomena determine the evolution and dynamics of wear and friction parameters at elevated subsurface temperatures [8-13]. Both individually and in the aggregate these factors may dramatically change the wear and friction evolution of the solid body especially if coexisting within the same temperature and contact stress ranges. However, there is a phenomenon that is manifested on a majority of materials at elevated temperatures. It is thermal expansion.

Control of thermal expansion is among the most important fundamental problems in friction physics but can be difficult to achieve. Actually, the majority of both natural and artificial materials possess a positive thermal expansion coefficient (TEC), whose magnitudes vary in the range $\alpha \sim 10^{-6}\text{-}10^{-5}$ K$^{-1}$. The positive thermal expansion (PTE) is a critical issue in many applications when a material experiences intense heating – aerospace engineering, electronics, etc. [14]. It is an especially important factor for extremely localized deformation and energy dissipation, i.e. in sliding friction conditions when two rough surfaces are contacting each other only by real contact areas and the intense heating results in additional (thermal-induced) stress between contacting and thermally expanding asperities.

These thermal-induced stresses have at least two negative effects on the sliding friction. The first is that these stresses facilitate reaching the local critical stress values and thus promote either inelastic deformation or fracture of the asperity. Secondly, these stresses provide the effect of surface separation which, even if a negligible macroscopic value, serves for reducing the number of contact spots and stress concentration on those left. The latter also leads to a fracture of contacting asperities. An integral effect of PTE is a wear rate growth and feasibility of a thermoelastic instability, the most general and intriguing type of instabilities during friction [15,16]. This term has been proposed by Barber [17], who was first to indicate that the self-supporting process of frictional heating of the surface may result in the total annihilation of the initial asperities and uncontrolled growth of wear and friction force (catastrophic wear).

A traditional solution to this fundamental problem is related either with selecting optimum external factor values (load, speed, contact geometry, lubrication, etc.) in order to reduce the contact stress or with applying special materials with heat- and wear-resistant stable structures and phase composition that provide the stability of distribution of contact spots [10,18,19]. A promising alternative approach is applying new adaptive materials, which possess negative thermal expansion (NTE). Over the past decade, the materials were developed, which reveal structural adaptation and provide contraction under heating [20-24] as well as different NTE performance temperature intervals. Materials possessing wide enough NTE temperature intervals as well as high thermal expansion coefficient values, for instance, zirconium tungstate $ZrW_2O_8$ with $\alpha \approx -3 \cdot 10^{-5}$ K$^{-1}$ [14,20,25]), are the most promising for being used to stabilize friction and wear. It is suggested that using the NTE effect in sliding may reduce scatter of asperity heights due to thermal-induced contraction and, therefore, result in more uniform load distribution as well as wear rate reduction. This brings us to the feasibility of "wearless sliding friction", i.e. sliding friction characterized by very mild or even zero wear at rather high sliding speeds and normal load. It is understandable that a truly "wearless sliding friction" regime is possible only in case of zero adhesion between the surfaces.

In our work, we introduce a novel theoretical study of the unlubricated wear of rough surfaces of NTE materials. We focus on revealing conditions under which thermal-induced contraction of asperities makes it possible to almost completely suppress the wear. To carry out the study, we developed a microscale model of sliding friction of a model NTE material with an absolutely rigid counterbody. Such a radical problem formulation implies that the friction energy dissipation occurs exclusively in a subsurface layer of NTE material so that the NTE effect on thermoelastic instability as well as friction and wear dynamics could be unambiguously evaluated.

The rough surface of NTE material was modeled by a number of independently deformable asperities so that each of them was characterized by $Y$-position of the top and some parameter called here as effective contact surface area $S$ [Fig. 1].

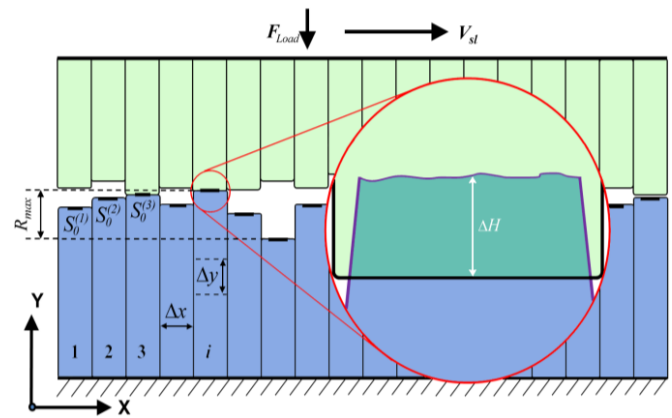

FIG. 1. Schematic diagram of sliding contact between rigid (top) and deformable (bottom) surfaces.

An instantaneous real contact area between asperities belonging to opposite surfaces is determined by plenty of factors and continuously changes during sliding [26-28]. Therefore, $S$-parameter could be defined as an effective parameter to characterize the time-average value of instantaneous real contact areas arising between two asperities belonging to opposite surfaces. For simplicity we assume that $S$ is the individual geometry characteristic of each deformable asperity, i.e. it is independent of the opposing contacting rigid asperity choice. Such a simplification has only negligible effect on the final results and corresponds to the case when the rigid body asperities are flat-top [Fig. 1]. Initial values of asperity's top position $Y_0^{(i)}$ and contact areas $S_0^{(i)}$, where $i$ is the asperity number, were drawn at random from homogeneous distributions at specified intervals $Y_0^{\min} \div Y_0^{\max}$ and $S_0^{\min} \div S_0^{\max}$, respectively. The roughness of the rigid countersurface was set similarly.

The surfaces were characterized by roughness parameters $R_{\max} = \left(Y_0^{\max} - Y_0^{\max}\right)$ and $R_a = 1/m \sum_{i=1}^{m} \left|Y_0^{(i)} - Y_{mean}\right|$, where $m$ is the number of asperities in the model, $Y_{mean} = 1/m \sum_{i=1}^{m} Y_0^{(i)}$ is the mean asperity height. The asperity width $\Delta x$ was equal for all the asperities of both surfaces.

Sliding friction interaction between deformable and rigid asperities involves both normal $F_N$ and tangential $F_\tau$ forces. Normal force $F_N$ provides compression of the deformable asperity by the rigid one and for simplicity reasons, it was determined using the spring stiffness $K$ approximation:

$$F_N = K\Delta H = K_{\min} \frac{S_0}{S_0^{\min}} \Delta H , \quad (1)$$

where $\Delta H = Y - Y_0$ is the asperity height reduction by compression, $K_{\min}$ is the stiffness of the minimum contact area $S_0^{\min}$ asperity. Here, we assume that asperity stiffness $K$ depends only on the asperity contact area. It is known from the solution of the problem of semi-infinite plate edge loaded by a point force that both elastic stresses and strains slowly fall off with the distance away from the force application point. The effect of the asperity height on the compression resistance is, therefore, assumed as negligible and ignored.

The tangential asperity interaction force was determined as $F_\tau = \mu F_N$ where $\mu$ is the local coefficient of friction kept constant for all contacts. Assuming that almost all mechanical energy dissipated during sliding friction transforms into heat, friction heat release power in the contact of two asperities was calculated from the friction force $F_\tau$ and sliding speed $V_{sl}$: $q = \mu F_N V_{sl}$. Heat removal from each asperity to the bulk of the deformable body was calculated by means of numerical solving the one-dimensional heat conductivity problem:

$$\frac{\partial T}{\partial t} = a \frac{\partial^2 T}{\partial y^2} , \quad (2)$$

where $a = k/(c\rho)$ is the thermal diffusivity, $k$ is the heat conductivity, $c$ is the specific heat capacity, $\rho$ is density. A heat source $q$ and constant temperature $T = T_0$ were used as boundary conditions on opposite ends of the asperity. A solution to the heat conductivity equation was obtained using the finite difference method with a step $\Delta y$ numerically equal to asperity width $\Delta x$. Here, we assume that the temperature of the asperity is associated with the temperature of the top cell. Heat transfer along the $X$-axis was ignored.

Thermal expansion leads to changing the surface roughness topography. It was assumed that since deeper surface layers are under constrain conditions and deform consistently then the temperature-induced change of the asperity height ($Y$-coordinate of the top of asperity) is almost totally determined by thermal expansion of the top cell:

$$Y = Y_0 + \alpha \Delta T \Delta y , \quad (3)$$

where $\Delta T = T - T_0$ is the asperity temperature difference between its current ($T$) and initial ($T_0$) values, $\alpha$ is the thermal expansion coefficient. Thermal expansion of the asperity also results in changing the effective contact area as follows:

$$S = S_0 (1 + \alpha \Delta T)^2 . \quad (4)$$

The value of equivalent stress in asperity of the deformable body was used as a criterion of asperity wear. The meaning of such a criterion is similar to that of commonly used von Mises stress criterion:

$$\sigma_{eq} = \frac{1}{\sqrt{2}} \sqrt{2\sigma_N^2 + 6\sigma_\tau^2} = \frac{F_N}{S} \sqrt{1 + 3\mu^2} \geq C , \quad (5)$$

where $\sigma_N = F_N/S$, $\sigma_\tau = F_\tau/S$, $C$ is the material strength characteristic. When inequality in the right part of (5) becomes true for some asperity, i.e. the criterion is fulfilled, then its height ($Y$-coordinate of the top) is reduced by quantity $dH_w$. Corresponding wear volume may be estimated as $dW_{asp} = dH_w S_0$. We assume that the asperity wear act does not change the corresponding values of $S_0$, $T$, and $K$.

The "elementary" wear volume $dH_w$ can vary widely since it depends on the majority of factors including wear mechanisms: mild wear, pulling out, fragmentation, etc. [29-33]. Generally, wear mechanism is determined by the contact pressure. Consequently, the worn material volume per elementary wear act $H_w$ is determined both by the ductility/brittleness ratio of the material and the applied stress level. Here, we assume simple linear dependence of the elementary wear $dH_w$ on equivalent stress $\sigma_{eq}$ achieved on the contact area:

$$H_w = H_w^{\min} + \frac{dH_w}{d\sigma_{eq}} (\sigma_{eq} - C), \quad (6)$$

where $dH_w/d\sigma_{eq} = const$, $H_w^{\min}$ is the asperity minimum wear volume at $\sigma_{eq} = C$.

Sliding friction was modeled by discretely moving the rigid counterbody over the surface of a damageable one along the $X$-axis at a constant sliding speed $V_{sl}$ [Fig. 1]. The time interval between two successive displacement acts $\Delta t = \Delta x/V_{sl}$ corresponded to the contact dwell time when frictional heat was released. The displacement step size was equal to the asperity width $\Delta x$ so that the whole system of asperity contacts was fully modified. No dynamic sliding effects including elastic wave propagation on the surface and in the bulk of the bodies were considered within this model. It was assumed that the system is under force equilibrium condition during each moment of time (it is a so-called "over-damped" system): $F_{Load} = \sum_{i=1}^{m} F_N^{(i)}$, where $F_{Load}$ is applied normal force, $F_N^{(i)}$ is the reaction force (1) from the $i$ asperity. Following this assumption, we determine the equilibrium $Y$-position of the rigid body surface with respect to deformable one at each time interval between successive displacement acts. The total wear was characterized using $W(t)$ dependencies, where $W = \sum_{t=0}^{n\Delta t} \sum_{i=1}^{m} H_w^{(i)} S_0^{(i)}$ is the wear debris volume, $n$ is current time step number. It is assumed, that wear debris are fully removed from the worn surface and thus have no effect on further course of surface interaction. The sliding friction process stability was estimated from analyzing the $W(t)$ dependencies for different contact stress levels. It is reasonable that steady sliding may be characterized by close to con-stant mild wear rate estimated as the mean value of $w = dW/dt$ derivative. On the contrary, instable (catastrophic) wear would be characterized by $w>0$. Also, the time stability\instability of the wear process would be imprinted on the worn surface topography.

We considered a deformable body model material, which possesses negative thermal expansion: thermal expansion coefficient $\alpha = -8.5 \cdot 10^{-5}$ K$^{-1}$, density $\rho = 9000$ kg/m$^3$, heat capacity c=380 J/(kg·K), heat conductivity $k = 5$ W/(m·K), Young's modulus $E = 125$ GPa, the strength of asperity material $C = 4$ GPa. The damageable and rigid surfaces were characterized by roughness $R_{max} = 2$ μm, $R_a \approx 0.5$ μm and $R_{max}^{rigid} = 0.5$ μm, $R_a^{rigid} \approx 0.125$ μm, respectively. The following parameters of asperity friction and wear were used: minimum elementary wear volume $H_w^{min} = 0.4$ μm, $dH_w/d\sigma_{eq} = 10^{-16}$ m/Pa, local coefficient of friction $\mu = 0.1$. The shown above values correspond to the characteristics of ceramic materials (including promising NTE ceramic materials) and typical surfaces in order of magnitude.

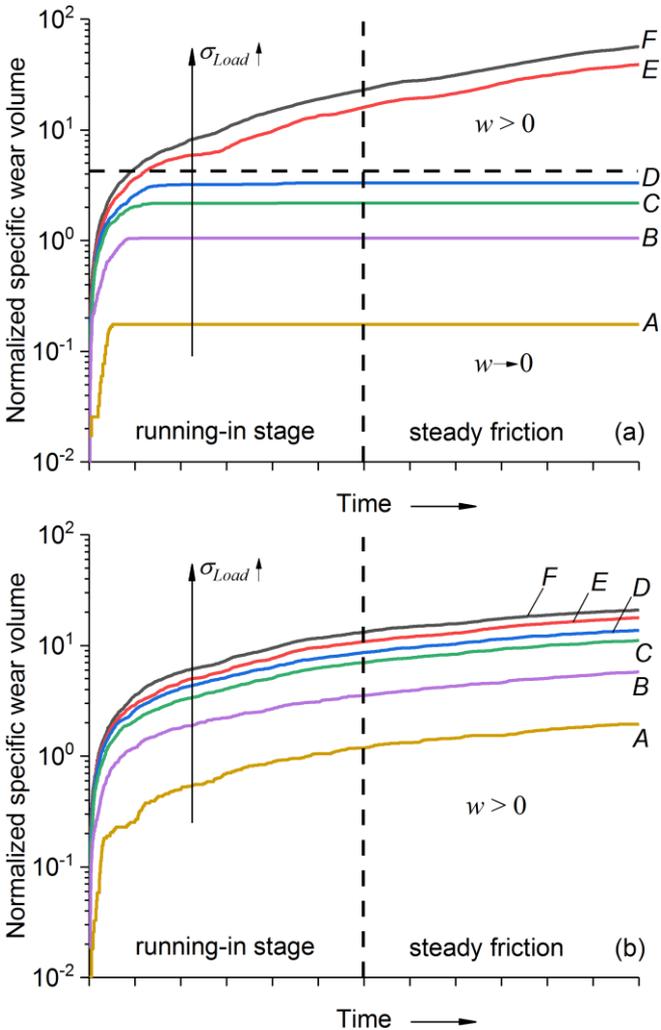

FIG. 2. Wear vs. time curves for NTE (a) and PTE (b) model materials tested at different applied normal stress levels $\sigma_{Load}/\sigma_{crit}$: 0.3 (A), 0.6 (B), 0.9 (C), 0.98 (D), 1.1 (E), 1.2 (F). The worn volume is shown here in dimensionless specific (per asperity) units $W' = W/(mH_w^{min} S_0^{max})$. Horizontal dashed line is a boundary between low wear and catastrophic wear stages. Vertical dashed line conditionally separates the running-in and steady friction stages.

Intense heating and wear of asperities is accompanied by establishing the constant levels of mean wear rate $w$ and subsurface temperature. The absolute value of $w$ is determined by the number of contact spots on which critical contact pressure $\sigma_N \geq C/\sqrt{1+3\mu^2}$ is achieved. The applied normal stress $\sigma_{Load} = F_{Load}/(m \cdot \Delta x^2)$ is, therefore, a key factor to determine the wear rate under tangential contacting.

The main fundamental result obtained on the NTE model material is the threshold nature of the $w(\sigma_{Load})$ dependency, i.e. there is some threshold value ($\sigma_{crit}$) of applied normal stress $\sigma_{Load}$ above which there occurs a transition from steady low wear ($w \to 0$) to catastrophic ($w>0$) wear. The wear evolution on NTE model material is represented in Fig. 2(a) as $W(t)$ dependencies plotted for different applied normal stresses. All curves demonstrate a short-time running-in stage when the wear rate decreases from the maximum initial value to a low steady value determined by $\sigma_{Load}$. The $W(t)$ curves may be divided into two sets. The first set was obtained at $\sigma_{Load}<\sigma_{crit}$ and these curves demonstrate a steady low wear regime of sliding (or "wearless friction"). The second set of curves relate to $\sigma_{Load}>\sigma_{crit}$ and their running-in stages give way to the increasing wear stage, which is commonly observed in dry friction on the PTE materials in the absence of any wear reduction mechanisms. Indeed, a similar study for PTE material with the same physical properties but TEC of the opposite sign ($\alpha = 8,5 \cdot 10^{-5}$) showed the absence of the "wearless friction" stage [Fig. 2(b)]. The steady wear stage is observed for all $\sigma_{Load}>0$ so that the mean wear rate is monotonically increased from mild to catastrophic level with the $\sigma_{Load}$.

Initially, the damageable surface is rather rough [Fig. 3(a)] and it becomes smoother in the process of sliding. The steady low wear friction for the model NTE material is characterized by one order of magnitude lower roughness of the worn surface [Fig. 3(b)]. Qualitatively different two-scale topography of the worn surface is observed for the catastrophic wear regime [Fig. 3(c)]. The main corrugation is characterized by periodic stepped profile (0.467 μm mean height and 178 μm mean width in the present case). The profile steps appear by means of successive shaving the deformable surface asperities by those of the rigid counter-surface. Such a shaving provides wear rate even higher than that of achieved on a PTE at the same values of $\sigma_{Load}$ [Fig. 2]. A lower order of magnitude roughness may be discovered on the profile step surfaces, whose parameters are comparable with those obtained at the steady low wear stage. The PTE model material worn surfaces demonstrate only one type of topography whose roughness slowly increases with the applied contact pressure, for example twice increase in applied normal stress leads to only 10% increase in $R_a$ [Fig. 3(d),(e)].

The difference between the two worn surface topographies obtained on the NTE material [Fig. 3(b),(c)] is explained by the temperature-induced shrinking of the asperity dimensions including height and effective contact area. Since surfaces are rough then local contact stresses $\sigma_N^{(i)}$ are distributed inhomogeneously and the most heavily loaded contact areas experience the highest frictional heating i.e. maximum thermal shrinking. The roughness is thus reduced and load is redistributed for previously less loaded asperities so that the maximum contact stresses reduce too and stress distribution becomes more homogeneous.

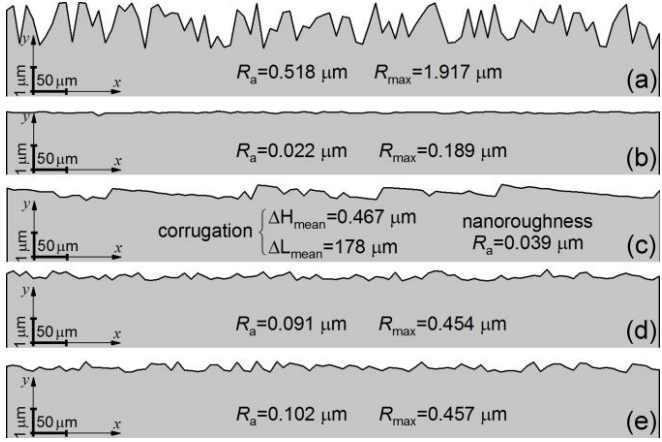

FIG. 3. The surface topography types found on the initial (a), and worn surfaces on NTE (b),(c) and PTE (d),(e) model materials. Here, (b) corresponds to NTE steady low wear mode at $\sigma_{Load}<\sigma_{crit}$; (c) corresponds to NTE catastrophic wear mode at $\sigma_{Load}>\sigma_{crit}$; (d) and (e) correspond to PTE catastrophic wear mode at $\sigma_{Load}=0.6\sigma_{crit}$ and $\sigma_{Load}=1.4\sigma_{crit}$, respectively. See Supplemental Material videos at [URL] for understanding the dynamics of roughness change.

The contact stress distribution histograms in Fig. 4(a),(b) show that when $\sigma_{Load} < \sigma_{crit}$, the maximum stresses stay below the asperity material strength $C$ and this is the condition for steady low wear sliding with only sporadic and local wear peaks originating from the sporadic nature of asperity height distribution. These peaks provide low but finite value of wear rate, which however stays close to zero.

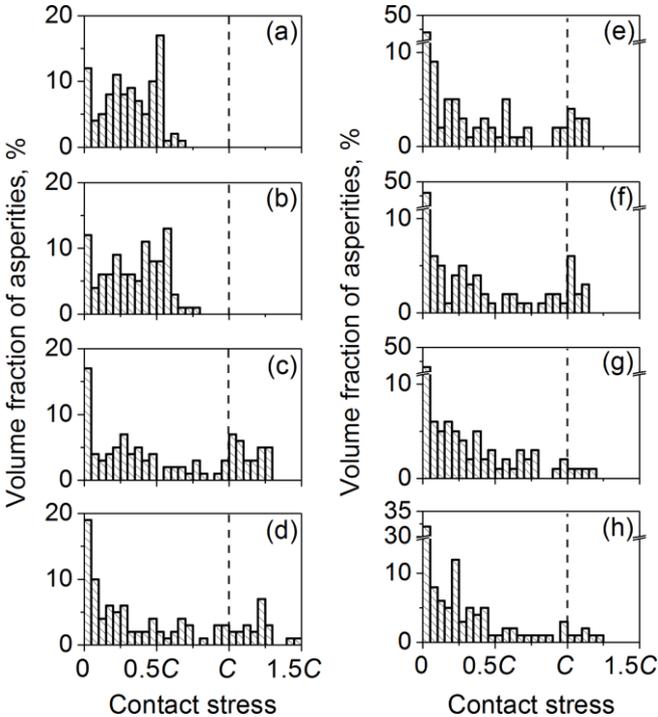

FIG. 4. Contact stress distribution histograms for NTE (a)-(d) and PTE (e)-(h) model materials at different applied contact pressure level $\sigma_{Load}$: (a),(e) $0.9\sigma_{crit}$, (b),(f) $0.98\sigma_{crit}$, (c),(g) $1.1\sigma_{crit}$, (d),(h) $1.2\sigma_{crit}$.

When $\sigma_{Load} > \sigma_{crit}$ there is always a fraction of asperities with the contact stresses above the ultimate stress $C$ [Fig. 4(c),(d)]. The higher is the applied contact pressure, the higher is the asperity fraction with the contact stress at the level of the material's strength. This may be a rationale behind the ever-increasing wear and specific type of the worn surface topography [Fig. 3(c)]. It was shown by means of parametric study that both the height and width of the corrugation are unambiguously related to the applied normal stress $\sigma_{Load}$. The higher is $\sigma_{Load}$, the higher is the fraction of rigid asperities that dig into the opposing surface and then cut shorter chips thus leaving the shorter marks on it.

The histograms in Fig. 4(e)-(h) for PTE materials testify that for all $\sigma_{Load}>0$ there is always a fraction of asperities with contact stresses above the material strength. This results from thermal elongation of the PTE asperities, which facilitates stress concentration on the highest ones and their ensuing wear even at low applied pressure levels.

So, we showed that frictional heating of the surface of NTE materials accompanied by temperature-induced smoothing and reduction of local contact pressures makes possible the implementation of steady low wear or even wearless friction at quite high applied normal loads. Such behavior is qualitatively different from that of PTE materials and may be related to a special type adaptation mechanism. The wear reduction effect of NTE makes the NTE materials extremely attractive for applications in both low- and heavy-loaded tribological devices, for example as components of composite friction (brake) materials. Controlling both mechanical and thermophysical characteristics of frictional materials, it is feasible to develop effective low or even zero wear rate tribocoupling.

The threshold stress $\sigma_{crit}$, which corresponds to a transition from low to catastrophic wear in sliding on NTE material, is determined by a number of material's thermal, physical, and mechanical characteristics. Mechanical behavior of the modeled body's surface is first of all defined by Young's modulus $E$ and ultimate strength $C$. Young's modulus determines the value of asperity stiffness $K \sim ES_0/D$ ($D$ is normalizing constant having length dimension) and defines the value of asperity strain in response to the applied contact stress. Reducing Young's modulus value results in increasing the number of contact spots and thus reducing the maximum local contact stress. Increasing the $E$ value i.e. increasing the deformable surface's stiffness provides the inverse effect. The ultimate stress $C$ determines the ultimate strain of the asperity, the maximum normal load, which the surface can sustain without fracture of asperities, and finally the time-averaged number of contacts during friction. Consequently, threshold stress $\sigma_{crit}$ should inversely depend on Young's modulus and directly depend on $C$. The simulation results show $\sigma_{crit}$ is inversely proportional to $E$ and directly proportional to $C^2$. Therefore, it is possible now defining a dimensionless combination $M = \sigma_{Load}E/C^2$, which unambiguously allows determining the threshold position between mild and catastrophic wear for NTE materials with different combinations of $E$ and $Y$ and other material characteristics and loading rate kept equal. Low steady and catastrophic wear will be thus realized at $M<M_{crit}$ and $M>M_{crit}$, respectively.

The particular value of $M_{crit}$ depends on the wear model parameters, i.e. on both $H_w^{min}$ and $dH_w/d\sigma_{eq}$. It was shown in the modeling when keeping the $dH_w/d\sigma_{eq}$ quantity constant that $M_{crit}(H_w^{min})$ dependence is a decreasing one and may be adequately approximated using a linear function as follows: $M_{crit} = M_{crit}^0 - AH_w^{min}$. The characteristic scale of $M_{crit}^0$ is of the order of unity. Dependence $M_{crit}(dH_w/d\sigma_{eq})$ is also a decreasing function.

Thermophysical properties such as heat conductivity $k$ and heat capacity $c$ are very important for sliding dynamics on the NTE materials since all the above-described wear specifics are deter-mined by an interplay between the heat release and removal processes. The relationship between *k* and *c* defines the asperity temperature dynamics at all wear stages thusly having its effect on the worn surface topography. Simulation results showed that increasing the heat conductivity by a factor of 10 provides more efficient heat removal and reducing the mean surface temperature by a factor of 1.5, heat conductivity kept constant. The reduced mean surface temperature leads to reducing the smoothing effect from the negative TEC. The threshold value $M_{crit}$ is decreased by 5-10%. Twofold increase/reduction of the heat capacity value provides increasing/reducing the mean surface temperature by 15-20% while dimensionless threshold $M_{crit}$ value is varied only within 10% inversely as heat capacity and analogously to that of heat conductivity *k*.

Both mechanical and thermophysical properties of the deformable body play a great role in governing the stability of wear and friction on the NTE materials. However, as far as $M_{crit}$ is concerned it should be noted that thermophysical properties are less important impact factor as compared to that of *E* and *C*.

All the above-presented results were obtained using only one initial rigid counter-surface roughness value, namely $R_a^{rigid}$ =0.5 µm. To study the effect of initial roughness on the transition from steady low to catastrophic wear ($M_{crit}$) a number of experiments were carried out by varying it from zero to $R_a^{rigid}$ =0.5 µm. It was found out that the threshold parameter decreased exponentially when increasing the roughness and such dependence may be adequately approximated by a function as follows: $M_{crit} \sim Be^{-R_a^{rigid}/R_0}$, where *B* is the constant and $R_0$ is the normalization coefficient which depends on the deformable surface initial roughness $R_a$ and equals 0.2-0.3$R_a$. For $R_a^{rigid}$ >0.6-0.7$R_a$ the $M_{crit}$ is close to zero, i.e. no steady low wear regime is feasible. Such a result is provided by the fact that the number of contacts does not sufficiently increase at the running-in stage to decrease maximum values of contact stresses below the ultimate stress *C*. Therefore, the temperature-induced surface smoothing and steady low wear on NTE materials may be reduced to almost nothing unless the rigid abrading surface has low enough relative roughness.

*Corresponding author: grigoriev@ispms.ru
[1] D. Kuhlmann-Wilsdorf, Mat. Sci. Eng. **93**, 119 (1987).
[2] M. Kalin and J. Vizintin, Wear **249**, 172 (2001).
[3] E. H. Smith and R. D. Arnell, Tribol. Lett. **52**, 407 (2013).
[4] H. A. Abdel-Aal, in *Encyclopedia of Tribology*, edited by Q. J. Wang, Y. W. Chung (Springer, Boston, MA, 2013).
[5] A. S. Iquebal, D. Sagapuram and S. T. S. Bukkapatnam, Sci. Rep. **9**, 10617 (2019).
[6] S. K. Lee, R. Han, E. J. Kim, G. Y. Jeong, H. Khim and T. Hirose, Nature Geosci. **10**, 436 (2017).
[7] M. Umar, R. A. Mufti and M. Khurram, Tribol. Int. **118**, 170 (2018).
[8] K. B. Jinesh, S. Yu. Krylov, H. Valk, M. Dienwiebel, and J. W. M. Frenken, Phys. Rev. B **78**, 155440 (2008).
[9] S. Yu. Krylov, K. B. Jinesh, H. Valk, M. Dienwiebel, and J. W. M. Frenken, Phys. Rev. E **71**, 065101(R) (2005).
[10] H. Spikes, Friction **6**, 1 (2018).
[11] N. L. Savchenko, Yu. A. Mirovoy, A. S. Buyakov, A. G. Burlachenko, M. A. Rudmin, I. N. Sevostyanova, S. P. Buyakova and S. Yu. Tarasov, Wear **446-447**, 203204 (2020).
[12] A. A. Voevodin, C. Muratore and S. M. Aouadi, Surf. Coat. Technol. **257**, 247 (2014).
[13] G. Cui, Y. Liu, S. Li, H. Liu, G. Gao and Z. Kou, Sci. Rep. **10**, 6816 (2020).
[14] K. Takenaka, Front. Chem. (Lausanne, Switz.) **6**, 267 (2018).
[15] S. Abbasi, S. Teimourimanesh, T. Vernersson, U. Sellgren, U. Olofsson, and R. Lundén, Wear **314**(1-2), 171 (2014).
[16] J. R. Barber, Proc. R. Soc. A **312**, 381 (1969).
[17] J. R. Barber, Wear **10**, 155 (1967).
[18] Y. Meng, J. Xu, Z. Jin, B. Prakash and Y. Hu, Friction **8**, 221 (2020).
[19] O. Hod, E. Meyer, Q. Zheng and M. Urbakh, Nature **563**, 485 (2018).
[20] J. Chen, L. Hu, J. Deng and X. Xing, Chem. Soc. Rev. **44**, 3522 (2015).
[21] A. G. Akulichev, B. Alcock, A. Tiwari and A. T. Echtermeyer, J. Mater. Sci. **51**, 10714 (2016).
[22] J. Ouyang, Y. Li, B. Chen and D. Huang, Materials **11**, 748 (2018).
[23] N. C. Burtch, S. J. Baxter, J. Heinen, A. Bird, A. Schneemann, D. Dubbeldam and A. P. Wilkinson, Adv. Funct. Mater. **48**, 1904669 (2019).
[24] J. Chen, Q. Gao, A. Sanson, X. Jiang, Q. Huang et al., Nat. Commun. **8**, 14441 (2017).
[25] T. A. Mary, J. S. O. Evans, T. Vogt, A. W. Sleight, Science **272**, 90 (1996).
[26] A. R. Hinkle, W. G. Nöhring, R. Leute, T. Junge, L. Pastewka, Sci. Adv. **6**, eaax0847 (2020).
[27] B. Weber, T. Suhina, T. Jungle, L. Pastewka, A. M. Brouwer and D. Bonn, Nat. Commun. **9**, 888 (2018).
[28] B. N. J. Persson, O. Albohr, U. Tartaglino, A. I. Volokitin, E. Tosatti, J. Phys. Condens. Matter **17**, R1 (2005).
[29] K. Kato, Wear **241**(2), 151 (2000).
[30] A. V. Dimaki, E. V. Shilko, I. V. Dudkin, S. G. Psakhie and V. L. Popov, Sci. Rep. **10**, 1585 (2020).
[31] R. Aghababaei, D. Warner and J. Molinari, Nat. Commun. **7**, 11816 (2016).
[32] R. Aghababaei, T. Brink and J. F. Molinari, Phys. Rev. Lett. **120**, 186105 (2018).
[33] J. von Lautz, L. Pastewka, P. Gumbsch, M. Moseler, Tribol. Lett. **63**, 26 (2016).